\documentclass[
notoc,nohyper]{JHEP} 

\usepackage[psamsfonts]{amsfonts}

\newcommand{\tr}{\mathop{\rm tr}\nolimits}
\newcommand{\str}{\mathop{\rm str}\nolimits}

\newcommand{\SU}{\mathop{\rm SU}}
\newcommand{\U}{\mathop{\rm {}U}}
\newcommand{\rmd}{{\rm d}}

\newcommand\fverb{\setbox\pippobox=\hbox\bgroup\verb}
\newcommand\fverbdo{\egroup\medskip\noindent%
                        \fbox{\unhbox\pippobox}\ }
\newcommand\fverbit{\egroup\item[\fbox{\unhbox\pippobox}]}
\newbox\pippobox

\title{
Chiral anomalies in the reduced model}

\author{Yoshio Kikukawa\\
Department of Physics, Nagoya University, Nagoya 464-8602, Japan\\
E-mail: \email{kikukawa@eken.phys.nagoya-u.ac.jp}}
\author{Hiroshi Suzuki\\
Department of Mathematical Sciences, Ibaraki University, Mito 310-8512, Japan\\
E-mail: \email{hsuzuki@mito.ipc.ibaraki.ac.jp}}
\received{\today}               
\accepted{\today}               

\preprint{DPNU-02-22\\IU-MSTP/51\\\heplat{0207009}}     

\abstract{
On the basis of an observation due to Kiskis, Narayanan and Neuberger, we show
that there is a remnant of chiral anomalies in the reduced model when a Dirac
operator which obeys the Ginsparg-Wilson relation is employed for the fermion
sector. We consider fermions belonging to the fundamental representation of the
gauge group~$\U(N)$ or~$\SU(N)$. For vector-like theories, we determine a
general form of the axial anomaly or the topological charge within a framework
of a $\U(1)$ embedding. For chiral gauge theories with the gauge group $\U(N)$,
a remnant of gauge anomaly emerges as an obstruction to a smooth fermion
integration measure. The pure gauge action of gauge-field configurations which
cause these non-trivial phenomena always diverges in the 't~Hooft $N\to\infty$
limit when~$d>2$.
}
\keywords{Renormalization Regularization and Renormalons, Lattice Gauge Field
Theories, Gauge Symmetry, Anomalies in Field and String Theories}

\begin{document} 

\maketitle 

\section{Introduction}
In a recent paper~\cite{Kiskis:2002gr}, Kiskis, Narayanan and Neuberger
proposed a use of the overlap-Dirac operator~\cite{Neuberger:1998fp} in the
quenched reduced model for the large~$N$
QCD~\cite{Eguchi:1982nm}--\cite{Das:1983pm} (for a more complete list of
references, see ref.~\cite{Das:1984nb}).\footnote{A similar proposal has been
made~\cite{Kitsunezaki:1997iu} in the context of the IIB matrix
model~\cite{Ishibashi:1996xs}.} In particular, they pointed out that it is
possible to define a topological charge~$Q$ in the reduced model in the spirit
of the overlap~\cite{Narayanan:1993wx,Randjbar-Daemi:1995sq}. Using the abelian
background of ref.~\cite{Giusti:2001ta}, they explicitly demonstrated that
certain configurations in the reduced model lead to~$Q\neq0$ for~$d=2$
and~$d=4$. They also argued that there may exist some remnant of the gauge
anomaly in reduced chiral gauge theories. These observations show an
interesting possibility that phenomena related to chiral anomalies in the
continuum gauge theory emerge even in the reduced model, although one would
naively expect there is no counterpart of chiral anomalies in the reduced model
in which spatial dependences of the gauge field are ``reduced''.

In this paper, we investigate this possibility further with a use of the
overlap- or a more general Dirac operator which obeys the Ginsparg-Wilson
relation~\cite{Ginsparg:1982bj,Hasenfratz:1998ft}. For our study, an exact
correspondence between the reduced model with restricted configurations and a
$\U(1)$ gauge theory defined on a finite-size lattice will be a basic tool. We
thus first clarify how to ``embed'' a $\U(1)$ lattice gauge theory in the
reduced model when fermion fields are belonging to the fundamental
representation of~$\U(N)$ or~$\SU(N)$ (section~2). Next, in section~3, after
characterizing the above topological charge~$Q$ as the axial anomaly in the
reduced model, we determine its general form within the $\U(1)$ embedding. For
this, a knowledge on the axial anomaly on finite-size
lattices~\cite{Igarashi:2002zz} is crucial; this knowledge is obtained by
combining cohomological analyses on the axial
anomaly~\cite{Luscher:1999kn}--\cite{Kikukawa:2001mw}, a complete
classification of ``admissible'' $\U(1)$ gauge
configurations~\cite{Luscher:1999du} and the locality of the Dirac
operator~\cite{Hernandez:1999et,Neuberger:2000pz}. We also show that, within
the $\U(1)$ embedding, the pure gauge action of any configuration with~$Q\neq0$
diverges in the 't~Hooft $N\to\infty$ limit; only exception is~$d=2$. In
section~4, we study reduced chiral gauge theories along the line of
refs.~\cite{Luscher:1999du,Luscher:2000un} and show that there exists an
obstruction to a smooth fermion integration measure over the space of
admissible reduced gauge fields; this obstruction might be regarded as a
remnant of the gauge anomaly. To show the obstruction, we utilize L\"uscher's
topological field in $d+2$-dimensional space~\cite{Luscher:2000un} and the
cohomological analysis applied to it~\cite{Kikukawa:2001kd}. Finally, in
section~5, we give a list of open questions and suggest directions of further
study.

\section{$\U(1)$ embedding}

In the most part of this paper, we focus only on the fermion sector and the
gauge field is treated as a non-dynamical background. In the reduced model, the
fermion action would be read as
\begin{equation}
   S_F=\overline\psi D\psi,
\label{twoxone}
\end{equation}
where $\psi$ and $\overline\psi$ are constant Grassman variables belonging to
the fundamental representation of~$\U(N)$ or~$\SU(N)$. The Dirac operator~$D$
defines a coupling of the fermion to the reduced gauge field~$U_\mu$. In the
case of the quenched reduced model~\cite{Bhanot:1982sh,Gross:at,Levine:1982uz},
the Dirac operator should be defined with a momentum insertion by the
factor~$e^{ip_\mu}$. As we will see below, such a global phase factor can be
absorbed into the $\U(1)$ gauge field within the $\U(1)$ embedding. So we will
omit the momentum factor in the following discussion.

The basic idea of an ``embedding'' is to identify the index~$n$
($1\leq n\leq N$) of the fundamental representation with the coordinate~$x$ on
a lattice with the size~$L$;
${\mit\Gamma}=\{\,x\in{\mathbb Z}^d\mid0\leq x_\mu<L\,\}$. We set $N=L^d$ and
adopt the convention between these two:
\begin{equation}
   n(x)=1+x_d+Lx_{d-1}+\cdots+L^{d-1}x_1,
\label{twoxtwo}
\end{equation}
where $x=(x_1,\ldots,x_d)\in{\mathbb Z}^d$. Note that $1\leq n(x)\leq L^d=N$.
With this mapping, a row vector $f_n$ is regarded as a function on the
lattice~$f(x)$; $f(x)=f_{n(x)}$. The shift operation on the
lattice\footnote{$\hat\mu$ denotes the unit vector in direction~$\mu$.}
\begin{equation}
   T_\mu^0f(x)=f(\widetilde{x+\hat\mu}),
\label{twoxthree}
\end{equation}
where $\widetilde x_\mu=x_\mu\bmod L$,
is then expressed by an action of the $N\times N$ matrix
\begin{equation}
   T_\mu^0=1\otimes\cdots\otimes1\otimes X\otimes1\otimes\cdots1,
\label{twoxfour}
\end{equation}
where the factor~$X$ appears in the $\mu$-th slot and each elements of the
tensor product are $L\times L$ matrices. The unitary matrix~$X$ is given by
\begin{equation}
   X=\pmatrix{0&1&&\cr
              & \ddots & \ddots &\cr
              &        & \ddots &1\cr
              1&&&0\cr}=VSV^\dagger,
\label{twoxfive}
\end{equation}
and
\begin{equation}
   S=\pmatrix{1&&&&\cr
                  &\eta&&&\cr
                  &&\eta^2&&\cr
                  &&&\ddots&\cr
                  &&&&\eta^{(L-1)}\cr},\qquad\eta=e^{2\pi i/L},
\label{twoxsix}
\end{equation}
because $X^L=1$.\footnote{Thus
$\det T_\mu^0=(\det S)^{L^{d-1}}=e^{\pi iL^{d-1}(L-1)}=1$ for~$d>1$.} In fact,
one verifies
\begin{equation}
   (T_\mu^0f)_{n(x)}=f_{n(\widetilde{x+\hat\mu})}=f(\widetilde{x+\hat\mu}).
\label{twoxseven}
\end{equation}
We may also define a {\it diagonal\/} $N\times N$ matrix from a function~$f(x)$
by
\begin{equation}
   f_{m(x)n(y)}=f_{n(x)}\delta_{m(x)n(y)}=f(x)\delta_{m(x)n(y)}.
\label{twoxeight}
\end{equation}
On this matrix, the shift is expressed by the conjugation
\begin{equation}
   (T_\mu^0fT_\mu^{0\dagger})_{m(x)n(y)}
   =f_{m(\widetilde{x+\hat\mu})n(\widetilde{y+\hat\mu})}
   =f(\widetilde{x+\hat\mu})\delta_{m(x)n(y)}.
\label{twoxnine}
\end{equation}

Now, the gauge coupling in the Dirac operator is always defined through the
covariant derivative. For the reduced model, the covariant derivative would be
read as
\begin{equation}
   \nabla_\mu\psi=U_\mu\psi-\psi.
\label{twoxten}
\end{equation}
We {\it assume\/} that the reduced gauge field $U_\mu$ has the following form
\begin{equation}
   U_\mu=u_\mu T_\mu^0,
\label{twoxeleven}
\end{equation}
with a {\it diagonal\/} matrix
\begin{equation}
   (u_\mu)_{m(x)n(y)}=(u_\mu)_{m(x)}\delta_{m(x)n(y)}
   =u_\mu(x)\delta_{m(x)n(y)}.
\label{twoxtwelve}
\end{equation}
Since $u_\mu$ is a unitary matrix,\footnote{When the gauge group is~$\SU(N)$,
we have an additional constraint that $\det u_\mu=1$ or
$\prod_{x\in{\mit\Gamma}}u_\mu(x)=1$.}
the diagonal elements are pure phase,
$(u_\mu)_{m(x)}=u_\mu(x)\in\U(1)$. We recall that in the conventional lattice
gauge theory the gauge coupling is defined through
\begin{eqnarray}
   \nabla_\mu\psi(x)&=&U_\mu(x)\psi(x+\hat\mu)-\psi(x)
\nonumber\\
   &=&U_\mu(x)T_\mu^0\psi(x)-\psi(x).
\label{twoxthirteen}
\end{eqnarray}
Comparing this with eqs.~(\ref{twoxten}) and~(\ref{twoxeleven}), we realize
that when the gauge field in the reduced model~$U_\mu$ has the particular
form~(\ref{twoxeleven}), the fermion sector in the reduced model is completely
identical to that of the conventional $\U(1)$ gauge theory defined on a lattice
with the size~$L$ ($N=L^d$). The $\U(1)$ link variables in the latter is given
by the diagonal elements of the $N\times N$ matrix~$u_\mu$. We call
eq.~(\ref{twoxeleven}) the $\U(1)$ embedding in this sense.

This identification has a gauge covariant meaning. Namely, the assumed
form~(\ref{twoxeleven}) is preserved under the gauge transformation in the
reduced model
\begin{equation}
   U_\mu\to\Omega U_\mu\Omega^\dagger,
\label{twoxfourteen}
\end{equation}
provided that $\Omega\in\U(N)$ or $\Omega\in\SU(N)$ is a {\it diagonal\/}
matrix. This transformation induces a transformation on $u_\mu$
\begin{equation}
   u_\mu\to\Omega u_\mu(T_\mu^0\Omega^\dagger T_\mu^{0\dagger}),
\label{twoxfifteen}
\end{equation}
that is nothing but the conventional $\U(1)$~gauge transformation due to
eq.~(\ref{twoxnine}).

Also the plaquette variable in the reduced model and that of the $\U(1)$ theory
have a simple relation under eq.~(\ref{twoxeleven}). We note\footnote{Note that
$[T_\mu^{0\dagger},T_\nu^0]=0$.}
\begin{equation}
   U_{\mu\nu}=U_\mu U_\nu U_\mu^\dagger U_\nu^\dagger
   =u_\mu(T_\mu^0u_\nu T_\mu^{0\dagger})(T_\nu^0u_\mu^\dagger T_\nu^{0\dagger})
   u_\nu^\dagger,
\label{twoxsixteen}
\end{equation}
is a diagonal matrix and the diagonal ($m(x)m(x)$) element of this equation is
the $\U(1)$ plaquette:
\begin{eqnarray}
   (U_{\mu\nu})_{m(x)m(x)}&=&
   (u_\mu)_{m(x)m(x)}(u_\nu)_{m(x+\hat\mu)m(x+\hat\mu)}
   (u_\mu)_{m(x+\hat\nu)m(x+\hat\nu)}^*(u_\nu)_{m(x)m(x)}^*
\nonumber\\
   &=&u_{\mu\nu}(x),
\label{twoxseventeen}
\end{eqnarray}
{}from eq.~(\ref{twoxnine}).

In the following, we utilize the above equivalence of the $\U(N)$ or $\SU(N)$
reduced model with restricted configurations and a $\U(1)$ gauge theory defined
on the finite lattice~${\mit\Gamma}$. Fortunately, when a Dirac operator which
obeys the Ginsparg-Wilson relation is employed, we may invoke a cohomological
analysis and related techniques which tell a structure of chiral anomalies on
a lattice with finite lattice spacings~\cite{Luscher:1999kn,Fujiwara:2000fi,%
Suzuki:2000ii,Kikukawa:2001kd,Luscher:2000zd,Kikukawa:2001mw} and with finite
sizes~\cite{Igarashi:2002zz}. We will fully use these powerful machineries to
investigate possible chiral anomalies in the reduced model.

\section{Axial anomaly and the topological charge}

Consider the average over fermion variables in the reduced model
\begin{equation}
   \langle{\cal O}\rangle_{\rm F}
   =\int\rmd\psi\rmd\overline\psi\,{\cal O}\exp(-\overline\psi D\psi),
\label{threexone}
\end{equation}
where we assume that the Dirac operator obeys the Ginsparg-Wilson
relation~\cite{Ginsparg:1982bj}
\begin{equation}
   \gamma_{d+1}D+\gamma_{d+1}D=D\gamma_{d+1}D.
\label{threextwo}
\end{equation}
The simplest choice is the overlap-Dirac operator~\cite{Neuberger:1998fp}
\begin{equation}
   D=1-A(A^\dagger A)^{-1/2},\qquad A=1-D_{\rm w},
\label{threexthree}
\end{equation}
where $D_{\rm w}$ is the standard Wilson-Dirac operator
\begin{equation}
   D_{\rm w}={1\over2}
   [\gamma_\mu(\nabla_\mu^*+\nabla_\mu)-\nabla_\mu^*\nabla_\mu].
\label{threexfour}
\end{equation}
The covariant derivative~$\nabla_\mu$ in the reduced model is defined by
eq.~(\ref{twoxten}) and $\nabla_\mu^*=\psi-U_\mu^\dagger\psi$. For the
overlap-Dirac operator to be well-defined, we require that the gauge field is
admissible~\cite{Hernandez:1999et,Neuberger:2000pz,Kiskis:2002gr}
\begin{equation}
   \|1-U_{\mu\nu}\|=\|1-U_\mu U_\nu U_\mu^\dagger U_\nu^\dagger\|<\epsilon,
\label{threexfive}
\end{equation}
where $\epsilon$ is a certain constant.

We make a change of variables in eq.~(\ref{threexone}),
$\psi\to\psi+\delta\psi$
and~$\overline\psi\to\overline\psi+\delta\overline\psi$,
where~\cite{Luscher:1998pq}
\begin{equation}
   \delta\psi=i\gamma_{d+1}\biggl(1-{1\over2}D\biggr)\psi,\qquad
   \delta\overline\psi=i\overline\psi\biggl(1-{1\over2}D\biggr)\gamma_{d+1}.
\label{threexsix}
\end{equation}
The fermion action does not change under this substitution due to the
Ginsparg-Wilson relation. The fermion measure however gives rise to a
non-trivial jacobian~$Q$ and we have
\begin{equation}
   \langle\delta{\cal O}\rangle_{\rm F}
   =2iQ\langle{\cal O}\rangle_{\rm F},\qquad
   Q=\tr\gamma_{d+1}\biggl(1-{1\over2}D\biggr).
\label{threexseven}
\end{equation}
We regard this jacobian as ``axial anomaly'' in the reduced model, because if
it were not present, a naive Ward-Takahashi identity
$\langle\delta{\cal O}\rangle_{\rm F}=0$ would be concluded from the symmetry
of the fermion action.

It is well-known that the combination~$Q$ is an
integer~\cite{Narayanan:1993wx,Hasenfratz:1998ri}. To see this, one notes that
the hermitian matrix~$\gamma_{d+1}D$ and $\gamma_{d+1}(1-D/2)$ anti-commute to
each other as a consequence of the Ginsparg-Wilson relation. If one evaluates
the trace in~$Q$ by using eigenfunctions of~$\gamma_{d+1}D$, therefore, only
zero-modes of~$\gamma_{d+1}D$ contribute; $Q$ is given by a sum of
$\gamma_{d+1}$~eigenvalues of zero-modes, i.e, the index. One may thus regard
$Q$ as the topological charge in the reduced model~\cite{Kiskis:2002gr}.

In general, it is not easy to write down $Q$ directly in terms of the reduced
gauge field~$U_\mu$. Nevertheless, at least for special configurations such
that
\begin{equation}
   U_\mu=\Omega u_\mu T_\mu^0\Omega^\dagger,
\label{threexeight}
\end{equation}
we can find the explicit form of $Q$ in terms of $U_\mu$ by using the
correspondence to a $\U(1)$ lattice gauge theory in the previous section. We
first note that the unitary matrix~$\Omega$ does not contribute to~$Q$, because
$Q$ is gauge invariant and $\Omega$ is the gauge transformation in the reduced
model. Then the gauge field has the form~(\ref{twoxeleven}). According to the
argument in the previous section, the system is completely identical to a
$\U(1)$ gauge theory. In particular, the trace in eq.~(\ref{threexseven}) is
replaced by the sum over all lattice sites. So we have
\begin{equation}
   Q=\sum_{x\in{\mit\Gamma}}\tr\gamma_{d+1}\biggl[1-{1\over2}D(x,x)\biggr],
\label{threexnine}
\end{equation}
where the $\U(1)$ gauge field is given by the diagonal elements of the
matrix~$u_\mu$. Note that the admissibility~(\ref{threexfive}) is promoted to
the admissibility in the $\U(1)$ theory, because $\|1-u_{\mu\nu}(x)\|<\epsilon$
for all $x$ from eq.~(\ref{twoxseventeen}) (recall that $U_{\mu\nu}$ is a
diagonal matrix).

Under the admissibility, a simple expression of $Q$~(\ref{threexnine}) in terms
of the $\U(1)$ gauge field is known. It is~\cite{Igarashi:2002zz}
\begin{eqnarray}
   Q&=&{(-1)^{d/2}\over(4\pi)^{d/2}(d/2)!}\sum_{x\in{\mit\Gamma}}
   \epsilon_{\mu_1\nu_1\cdots\mu_{d/2}\nu_{d/2}}
   f_{\mu_1\nu_1}(x)f_{\mu_2\nu_2}(x+\hat\mu_1+\hat\nu_1)\cdots
\nonumber\\
   &&\qquad\qquad\qquad\quad\times f_{\mu_{d/2}\nu_{d/2}}
   (x+\hat\mu_1+\hat\nu_1+\cdots+\hat\mu_{d/2-1}+\hat\nu_{d/2-1}),
\label{threexten}
\end{eqnarray}
where the $\U(1)$ field strength is defined by\footnote{For the cohomological
analysis to apply, $\epsilon$ in eq.~(\ref{threexfive}) has to be smaller than
$1$. Then the logarithm of the plaquette always remains within the principal
branch because $|f_{\mu\nu}(x)|<\pi/3$.}
\begin{equation}
   f_{\mu\nu}(x)={1\over i}\,\ln u_{\mu\nu}(x),\qquad-\pi<f_{\mu\nu}(x)\leq\pi.
\label{threexeleven}
\end{equation}
Thus, we immediately find, in the reduced model
\begin{eqnarray}
   Q&=&{i^{d/2}\over(4\pi)^{d/2}(d/2)!}\,
   \epsilon_{\mu_1\nu_1\cdots\mu_{d/2}\nu_{d/2}}
   \tr(\ln U_{\mu_1\nu_1})
    T_{\mu_1}^0T_{\nu_1}^0
   (\ln U_{\mu_2\nu_2})T_{\nu_1}^{0\dagger}T_{\mu_1}^{0\dagger}\cdots
\nonumber\\
   &&\qquad\times
   T_{\mu_1}^0T_{\nu_1}^0\cdots T_{\mu_{d/2-1}}^0T_{\nu_{d/2-1}}^0
   (\ln U_{\mu_{d/2}\nu_{d/2}})
   T_{\nu_{d/2-1}}^{0\dagger}T_{\mu_{d/2-1}}^{0\dagger}\cdots
   T_{\nu_1}^{0\dagger}T_{\mu_1}^{0\dagger}.
\label{threextwelve}
\end{eqnarray}
Note that $T_{\mu_1}^0T_{\nu_1}^0(\ln U_{\mu_2\nu_2})%
T_{\nu_1}^{0\dagger}T_{\mu_1}^{0\dagger}$ for example is Lie-algebra valued.
Since this is a diagonal matrix, it belongs to the Cartan sub-algebra.
Therefore, $Q$ is given by a linear combination
of~$\str(T^{a_1}\cdots T^{a_{d/2}})$, where $T^a$ is a (Cartan) generator of
the gauge group in the fundamental representation.

We want to evaluate $Q$ for admissible configurations. Fortunately, admissible
$\U(1)$ gauge fields have been completely classified by
L\"uscher~\cite{Luscher:1999du}. The most general form of the $\U(1)$ link
variable such that~$\|1-u_{\mu\nu}(x)\|<\epsilon$ for all~$x$ is given
by\footnote{When the gauge group is~$\SU(N)$,
$\prod_{x\in{\mit\Gamma}}u_\mu(x)$ must be unity. This requires that
$w_\mu\in{\mathbb Z}_{L^{d-1}}$ and $\prod_{x\in{\mit\Gamma}}v_\mu^{[m]}(x)=%
\exp[-\pi iL^{d-2}(L-1)\sum_\nu m_{\mu\nu}]=1$. The latter is always satisfied
for~$d>2$.}
\begin{equation}
   u_\mu(x)=\omega(x)v_\mu^{[m]}(x)u_\mu^{[w]}(x)
   e^{ia_\mu^{\rm T}}(x)\omega(x+\hat\mu)^{-1}.
\label{threexthirteen}
\end{equation}
In this expression, $\omega(x)\in\U(1)$ is the $\U(1)$ gauge transformation.
The field~$u_\mu^{[w]}(x)$ is defined by
\begin{equation}
   u_\mu^{[w]}(x)=\cases{w_\mu,&for $x_\mu=0$,\cr
                         1,&otherwise,\cr}\qquad w_\mu\in\U(1),
\label{threexfourteen}
\end{equation}
and it has vanishing field strength~$f_{\mu\nu}(x)=0$ and carries the Wilson
(or Polyakov) line, $\prod_{s=0}^{L-1}u_\mu^{[w]}(s\hat\mu)=w_\mu$. The
field~$v_\mu^{[m]}(x)$ is defined by
\begin{equation}
   v_\mu^{[m]}(x)=\exp\biggl[
   -{2\pi i\over L^2}\biggl(
   L\delta_{x_\mu,L-1}\sum_{\nu>\mu}m_{\mu\nu}x_\nu
   +\sum_{\nu<\mu}m_{\mu\nu}x_\nu
   \biggr)\biggr],
\label{threexfifteen}
\end{equation}
and carries a constant field strength
\begin{equation}
   f_{\mu\nu}(x)={2\pi\over L^2}\,m_{\mu\nu},
\label{threexsixteen}
\end{equation}
where the ``magnetic flux'' $m_{\mu\nu}$ is an integer bounded
by\footnote{$\epsilon'=2\arcsin(\epsilon/2)$.}
\begin{equation}
   |m_{\mu\nu}|<{\epsilon'\over2\pi}\,L^2.
\label{threexseventeen}
\end{equation}
The ``transverse'' gauge potential~$a_\mu^{\rm T}(x)$ is defined
by\footnote{$\partial_\mu$ and~$\partial_\mu^*$ denote the forward and the
backward difference operators, $\partial_\mu f(x)=f(x+\hat\mu)-f(x)$,
$\partial_\mu^*f(x)=f(x)-f(x-\hat\mu)$, respectively.}
\begin{eqnarray}
   &&\partial_\mu^*a_\mu^{\rm T}(x)=0,\qquad
   \sum_{x\in{\mit\Gamma}}a_\mu^{\rm T}(x)=0,
\nonumber\\
   &&|f_{\mu\nu}(x)|
   =|\partial_\mu a_\nu^{\rm T}(x)-\partial_\nu a_\mu^{\rm T}(x)
   +2\pi m_{\mu\nu}/L^2|<\epsilon'.
\label{threexeighteen}
\end{eqnarray}
Note that the space of~$a_\mu^{\rm T}(x)$ is contractible.

In terms of $N\times N$ matrix in the reduced model, the above admissible
configuration is represented by [$\omega(x)$ can be absorbed into $\Omega$
in eq.~(\ref{threexeight})]
\begin{equation}
   U_\mu=u_\mu T_\mu^0=v_\mu^{[m]}u_\mu^{[w]}e^{ia_\mu^{\rm T}}T_\mu^0,\qquad
   a_\mu^{\rm T}-T_\mu^{0\dagger}a_\mu^{\rm T}T_\mu^0=0,\qquad
   \tr a_\mu^{\rm T}=0,
\label{threexnineteen}
\end{equation}
where
\begin{equation}
   u_\mu^{[w]}=1\otimes\cdots\otimes1\otimes
   W_\mu\otimes1\otimes\cdots\otimes1,
\label{threextwenty}
\end{equation}
with
\begin{equation}
   W_\mu=\pmatrix{w_\mu&&&\cr
                  &1&&\cr
                  &&\ddots&\cr
                  &&&1\cr},\qquad w_\mu\in\U(1),
\label{threextwentyone}
\end{equation}
and
\begin{equation}
   v_\mu^{[m]}=Y^{m_{1\mu}}\otimes\cdots\otimes Y^{m_{\mu-1\mu}}\otimes Z_\mu,
\label{threextwentytwo}
\end{equation}
where
\begin{equation}
   Y=\pmatrix{1&&&&\cr
                  &\zeta&&&\cr
                  &&\zeta^2&&\cr
                  &&&\ddots&\cr
                  &&&&\zeta^{(L-1)}\cr},\qquad\zeta=e^{2\pi i/L^2},
\label{threextwentythree}
\end{equation}
and
\begin{eqnarray}
   Z_\mu&=&\pmatrix{1&&&\cr
                   &\ddots&&\cr
                   &      &1&\cr
                   &      & &0\cr}\otimes1\otimes\cdots\otimes1
\nonumber\\
   &&\qquad+\pmatrix{0&&&\cr
                   &\ddots&&\cr
                   &      &0&\cr
                   &      & &1\cr}
   \otimes S^{m_{\mu+1\mu}}\otimes\cdots\otimes
   S^{m_{d\mu}}.
\label{threextwentyfour}
\end{eqnarray}
For the configuration~(\ref{threexnineteen}) or equivalently for
eq.~(\ref{threexthirteen}), from eq.~(\ref{threexten}), we have
\begin{equation}
   Q={(-1)^{d/2}\over2^{d/2}(d/2)!}\,
   \epsilon_{\mu_1\nu_1\cdots\mu_{d/2}\nu_{d/2}}
   m_{\mu_1\nu_1}m_{\mu_2\nu_2}\cdots m_{\mu_{d/2}\nu_{d/2}},
\label{threextwentysix}
\end{equation}
which is manifestly an integer. This is the general form of the axial anomaly
in the reduced model within the $\U(1)$ embedding. We note that
$|Q|<{\epsilon'}^{d/2}d!L^d/[(4\pi)^{d/2}(d/2)!]\propto N$.

It is interesting to consider the pure gauge action
\begin{eqnarray}
   S_G&=&N\beta\sum_{\mu,\nu}\mathop{\rm Re}\tr(1-U_{\mu\nu})
\nonumber\\
   &=&N\beta\sum_{\mu,\nu}\sum_{x\in{\mit\Gamma}}[1-\cos f_{\mu\nu}(x)],
\label{threextwentyseven}
\end{eqnarray}
of an admissible configuration\footnote{To make the admissibility and a
smoothness of the action compatible, this action might be too
simple~\cite{Luscher:1999du}.} with~$Q\neq0$. For $u_\mu(x)=v_\mu^{[m]}(x)$,
this reads,
\begin{eqnarray}
   S_G&=&N\beta\sum_{\mu,\nu}\sum_{x\in{\mit\Gamma}}
   \biggl(1-\cos{2\pi\over L^2}m_{\mu\nu}\biggr)
\nonumber\\
   &\buildrel{N\to\infty}\over\rightarrow&
   2\pi^2\beta N^{2-4/d}\sum_{\mu,\nu}m_{\mu\nu}^2,
\label{threextwentyeight}
\end{eqnarray}
where we have used~$N=L^d$. Thus, as noted in ref.~\cite{Kiskis:2002gr}, the
action of~$u_\mu(x)=v_\mu^{[m]}(x)$ remains finite only for~$d=2$ (allowed
fluctuations of $a_\mu^{\rm T}(x)$ are of $O(1/N)$). In fact, this behavior
persists for general admissible configurations:
\begin{eqnarray}
   S_G&\geq&N\beta\sum_{\mu,\nu}\sum_{x\in{\mit\Gamma}}\alpha f_{\mu\nu}(x)^2
\nonumber\\
   &=&N\beta\alpha\sum_{\mu,\nu}\sum_{x\in{\mit\Gamma}}
   \biggl\{[\partial_\mu a_\nu^{\rm T}(x)-\partial_\nu a_\mu^{\rm T}(x)]^2
   +{4\pi^2\over L^4}m_{\mu\nu}^2\biggr\}
\nonumber\\
   &\geq&4\pi^2\alpha\beta N^{2-4/d}\sum_{\mu,\nu}m_{\mu\nu}^2,
\label{threextwentynine}
\end{eqnarray}
where, in the first line, we have noted $\cos x\leq1-\alpha x^2$
for~$0<\alpha<1/2$. This lower bound for the action shows that the action of a
configuration with $Q\neq0$ always diverges for $N\to\infty$ if $d>2$, within
the $\U(1)$ embedding.

\section{Obstruction to a smooth measure in reduced chiral gauge theories}

In this section, we consider a Weyl fermion coupled to the reduced gauge field
and show that there is an obstruction to a smooth fermion measure; this might
be regarded as a remnant of the gauge anomaly of the original theory.

The average over fermion variables is defined by\footnote{The
presentation in this section closely follows the framework
of~refs.~\cite{Luscher:1999du,Luscher:2000un}. We refer
to~refs.~\cite{Luscher:1999du,Luscher:2000un} and references therein for
further details.}
\begin{equation}
   \langle{\cal O}\rangle_{\rm F}
   =\int{\rm D}[\psi]{\rm D}[\overline\psi]\,
   {\cal O}\exp(-\overline\psi D\psi),
\label{fourxone}
\end{equation}
where Weyl fermions are subject of the chirality constraint
\begin{equation}
   \hat P_{\rm H}\psi=\psi,\qquad
   \overline\psi P_{\widetilde{\rm H}}=\overline\psi.
\label{fourxtwo}
\end{equation}
In this expression, the chiral projectors are defined by
$\hat P_\pm=(1\pm\hat\gamma_{d+1})/2$ and~$P_\pm=(1\pm\gamma_{d+1})/2$ and
$\hat\gamma_{d+1}$ is the modified chiral matrix,
$\hat\gamma_{d+1}=\gamma_{d+1}(1-D)$; ${\rm H}$ denotes the
chirality~${\rm H}=\pm$ and~$\widetilde{\rm H}=\mp$. Note that the
Ginsparg-Wilson relation implies $(\hat\gamma_{d+1})^2=1$ and
$D\hat\gamma_{d+1}=-\gamma_{d+1}D$. This definition thus provides a consistent
decomposition of the fermion action,
$\overline\psi D\hat P_{\rm H}\psi=\overline\psi P_{\widetilde{\rm H}}D\psi$.

The fermion integration measure is defined as usual
by~${\rm D}[\psi]=\prod_j\rmd c_j$, where $c_j$ is the expansion coefficient
in $\psi=\sum_jv_jc_j$ with respect to an orthonormal basis~$v_j$ in the
constrained space~$\hat P_{\rm H}v_j=v_j$
[$(v_k,v_j)=\delta_{kj}$].\footnote{For the anti-fermion,
${\rm D}[\overline\psi]=\prod_k\rmd\overline c_k$, where
$\overline\psi=\sum_k\overline c_k\overline v_k$ and
$\overline v_kP_{\widetilde{\rm H}}=\overline v_k$. Basis
vectors~$\overline v_k$ can be chosen to be independent of the gauge field.}
However, since the chiral projector~$\hat P_{\rm H}$ depends on the gauge
field, and the constraint~$\hat P_{\rm H}v_j=v_j$ alone does not specify basis
vectors uniquely, it is not obvious how one should change the basis
vectors~$v_j$ when the gauge field is varied. This implies that there exists a
gauge-field-depending phase ambiguity in the measure. This problem is
formulated as follows:

One can cover the space of admissible configurations by open local coordinate
patches~$X_A$ labelled by an index~$A$. Within each patch, smooth basis
vectors~$v_j^A$ can always be found, because $\hat P_{\rm H}$ depends smoothly
on the gauge field. In the intersection $X_A\cap X_B$, however, two bases are
in general different and related by a unitary transformation,
$v_j^B=\sum_lv_l^A\tau(A\to B)_{lj}$
and~$c_j^B=\sum_l\tau(A\to B)_{jl}^{-1}c_l^A$. The fermion measures defined
with respect to each basis are thus related as
\begin{equation}
   {\rm D}[\psi]^B=g_{AB}\,{\rm D}[\psi]^A,\qquad
   g_{AB}=\det\tau(A\to B)\in\U(1).
\label{fourxthree}
\end{equation}
Hence the above setup defines a $\U(1)$ fiber bundle over the space of
admissible configurations, $g_{AB}$ being the transition function. The
smoothness of the fermion integration measure (i.e., single-valued-ness of
$\langle{\cal O}\rangle_{\rm F}$) thus requires that this $\U(1)$ bundle is
trivial and that one can adjust bases~$v_j^A$ and $v_j^B$ such that the
transition function is unity, $g_{AB}=1$ on~$X_A\cap X_B$.\footnote{Under a
change of bases, the transition function transforms according to
$g_{AB}\to h_Ag_{AB}h_B^{-1}$ on~$X_A\cap X_B$, where $h_A$ ($h_B$) is a
determinant of the transformation matrix in the patch~$X_A$ ($X_B$).} Whether
this is the case or not eventually depends on the properties of the chiral
projector~$\hat P_{\rm H}$ and of the base manifold, the space of admissible
configurations.

We consider an infinitesimal variation of the gauge field
\begin{equation}
   \delta_\eta U_\mu=\eta_\mu U_\mu,\qquad\eta_\mu=\eta_\mu^aT^a,
\label{fourxfour}
\end{equation}
and introduce the ``measure term'' in the patch~$X_A$ by
\begin{equation}
   {\mathfrak L}_\eta^A=i\sum_j(v_j^A,\delta_\eta v_j^A),
\label{fourxfive}
\end{equation}
which parameterizes the above phase ambiguity. The measure terms in adjacent
two patches are related by
\begin{equation}
   {\mathfrak L}_\eta^A={\mathfrak L}_\eta^B-i\delta_\eta\ln g_{AB},\qquad
   {\rm on}\quad X_A\cap X_B.
\label{fourxsix}
\end{equation}
Thus the measure term is the connection of the $\U(1)$ bundle. We may introduce
a local coordinate $(t,s,\ldots)$ in~$X_A$ and define the $\U(1)$ curvature by
\begin{equation}
   \partial_t{\mathfrak L}_\sigma^A-\partial_s{\mathfrak L}_\tau^A
   =i\tr(\hat P_{\rm H}[\partial_t\hat P_{\rm H},\partial_s\hat P_{\rm H}]),
\label{fourxseven}
\end{equation}
where the variation vectors have been defined by
\begin{equation}
   \tau_\mu=\partial_tU_\mu U_\mu^\dagger,\qquad
   \sigma_\mu=\partial_sU_\mu U_\mu^\dagger.
\label{fourxeight}
\end{equation}
Equation~(\ref{fourxseven}), which follows from eq.~(\ref{fourxfive})
and~$[\partial_t,\partial_s]=0$, shows that the curvature is independent
of the referred patch, as it should be the case.\footnote{The above
$\U(1)$~bundle, the connection and the curvature were first addressed
in~ref.~\cite{Neuberger:1998xn} in the context of the overlap.}

Take a closed 2~dimensional surface~${\cal M}$ in the space of admissible
configurations. The first Chern number of the above $\U(1)$~bundle is then
given by
\begin{equation}
   {\cal I}={1\over2\pi}\int_{\cal M}{\rm d}t\,{\rm d}s\,
   i\tr(\hat P_{\rm H}[\partial_t\hat P_{\rm H},\partial_s\hat P_{\rm H}]).
\label{fourxnine}
\end{equation}
If this integer does not vanish, ${\cal I}\neq0$, the $\U(1)$~bundle is
non-trivial and a smooth fermion measure does not exist according to the above
argument. If ${\cal I}\neq0$, we may regard this as a remnant of the gauge
anomaly, because in the classical continuum limit of the original gauge theory
{\it before the reduction}, ${\cal I}$ is proportional to the
anomaly~$\str[R(T^{a_1})\cdots R(T^{a_{d/2+1}})]$, where $R$~is the gauge
representation of the Weyl
fermion~\cite{Luscher:2000un,Adams:2000yi}.\footnote{Under the infinitesimal
gauge transformation,
$\delta_\eta U_\mu=[\omega,U_\mu]$, $\delta_\eta\psi=\omega\psi$
and~$\delta_\eta\overline\psi=-\overline\psi\omega$, one can show that
\begin{eqnarray}
   &&\delta_\eta\langle{\cal O}\rangle_{\rm F}
   =\langle\delta_\eta{\cal O}\rangle_{\rm F}
   +i\omega^a[{\cal A}^a-(\nabla_\mu^*j_\mu)^a]\langle{\cal O}\rangle_{\rm F},
\nonumber\\
   &&\nabla_\mu^*j_\mu=j_\mu-U_\mu^\dagger j_\mu U_\mu,\qquad
   {\cal A}^a=-i\tr T^a\gamma_{d+1}\biggl(1-{1\over2}D\biggr),
\end{eqnarray}
where $j_\mu$ is the measure current defined
by~${\mathfrak L}_\eta=\eta_\mu^aj_\mu^a$, where $\eta_\mu=-\nabla_\mu\omega$
and~$\nabla_\mu\omega=U_\mu\omega U_\mu^\dagger-\omega$. The gauge anomaly in
this framework is thus given by the combination,
${\cal G}^a={\cal A}^a-(\nabla_\mu^*j_\mu)^a$. An evaluation of~${\cal G}^a$ is
however somewhat subtle because it is ambiguous depending on the measure
current which specifies the fermion integration measure. For conventional
chiral gauge theories, assuming the locality of the measure current, it is
possible to argue that this ambiguity can be absorbed into a gauge variation of
a local functional (i.e., a local counter-term). In the reduced model, however,
the meaning of the locality of the measure current~$j_\mu^a$ is not clear. This
is the reason why we study the first Chern number~${\cal I}$ instead of the
gauge anomaly~${\cal G}^a$ itself.}

The above is for the reduced model. The correspondence to the $\U(1)$ theory
in~section~2 is applied also to this system of Weyl fermion, because couplings
to the gauge field, even in the chiral constraint~(\ref{fourxtwo}), arise only
through the covariant derivative~(\ref{twoxten}). Hence, under the
assumption~(\ref{twoxeleven}), the above system is identical to a $\U(1)$
chiral gauge theory defined on the lattice~${\mit\Gamma}$ in which the
Ginsparg-Wilson Dirac operator is employed. In terms of the $\U(1)$ lattice
theory, the first Chern number reads
\begin{equation}
   {\cal I}={1\over2\pi}\int_{\cal M}{\rm d}t\,{\rm d}s\,
   i\sum_{x\in{\mit\Gamma}}
   \tr(\hat P_{\rm H}[\partial_t\hat P_{\rm H},\partial_s\hat P_{\rm H}])(x,x).
\label{fourxten}
\end{equation}
We will evaluate ${\cal I}$ in this $\U(1)$~picture. Since this ${\cal I}$ is
an integer, it is invariant under a smooth deformation of admissible
configurations defined on ${\cal M}$. This implies that ${\cal I}$ is
independent of the transverse potential~$a_\mu^{\rm T}(x)$
in~eq.~(\ref{threexthirteen}), because these degrees of freedom can be deformed
to the trivial value, $a_\mu^{\rm T}(z)\to0$, without affecting the
admissibility.

To evaluate~${\cal I}$ in the picture of $\U(1)$~lattice theory, it is
convenient to introduce L\"uscher's topological field in $d+2$-dimensional
space~\cite{Luscher:2000un}. To define this field, we introduce continuous
two dimensional space whose coordinates are $t$ and~$s$. The $\U(1)$ gauge
field is assumed to depend also on these additional coordinates, $u_\mu(z)$
where~$z=(x,t,s)$. We further introduce gauge potentials $a_t(z)$,
$a_s(z)\in{\mathfrak u}(1)$ along the continuous directions. The associated
field tensor is defined by
\begin{equation}
   f_{ts}(z)=\partial_ta_s(z)-\partial_sa_t(z),
\label{fourxeleven}
\end{equation}
and the covariant derivatives is defined by ($r=t$ or~$s$)
\begin{equation}
   D_r^au_\mu(z)=\partial_ru_\mu(z)+ia_r(z)u_\mu(z)-iu_\mu(z)a_r(z+\hat\mu).
\label{fourxtwelve}
\end{equation}
For a gauge covariant quantity such that~$\hat P_{\rm H}$, it reads
\begin{equation}
   D_r^a\hat P_{\rm H}=\partial_r\hat P_{\rm H}+i[a_r,\hat P_{\rm H}].
\label{fourxthirteen}
\end{equation}
L\"uscher's topological field is then defined by\footnote{$\epsilon_\pm=\pm1$.}
\begin{equation}
   q(z)=i\epsilon_{\rm H}\tr\biggl\{\biggl[
   {1\over4}\hat\gamma_{d+1}[D^a_t\hat P_{\rm H},D^a_s\hat P_{\rm H}]
   +{1\over4}[D^a_t\hat P_{\rm H},D^a_s\hat P_{\rm H}]\hat\gamma_{d+1}
   +{i\over2}f_{ts}\hat\gamma_{d+1}\biggr](x,x)\biggr\},
\label{fourxfourteen}
\end{equation}
which is a gauge invariant (in $d+2$-dimensional sense) pseudoscalar local
field. It can be verified that~\cite{Luscher:2000un}
\begin{equation}
   \sum_{x\in{\mit\Gamma}}q(z)
   =i\sum_{x\in{\mit\Gamma}}\tr\biggl[
   \hat P_{\rm H}[\partial_t\hat P_{\rm H},
   \partial_s\hat P_{\rm H}]
   +{i\over2}\epsilon_{\rm H}\partial_t(a_s\hat\gamma_{d+1})
   -{i\over2}\epsilon_{\rm H}\partial_s(a_t\hat\gamma_{d+1})\biggr](x,x).
\label{fourxfifteen}
\end{equation}
Thus it is a topological field satisfying
\begin{equation}
   \int{\rm d}t\,{\rm d}s\sum_{x\in{\mit\Gamma}}\delta q(z)=0,
\label{fourxsixteen}
\end{equation}
for any local variation of the gauge fields, $u_\mu(z)$ and~$a_r(z)$.
Equation~(\ref{fourxfifteen}) also shows that
\begin{equation}
   {\cal I}={1\over2\pi}\int_{\cal M}{\rm d}t\,{\rm d}s
   \sum_{x\in{\mit\Gamma}}q(z),
\label{fourxseventeen}
\end{equation}
if the gauge fields, $u_\mu(z)$ and~$a_r(z)$, are single-valued on~${\cal M}$.

A cohomological analysis again provides an important information on~$q(z)$.
Using the gauge invariance, the topological property and the pseudoscalar
nature of~$q(z)$, a cohomological analysis along the line
of~ref.~\cite{Kikukawa:2001kd} shows that
\begin{equation}
   q^\infty(z)=p(z)+\partial_\mu^*k_\mu^\infty(z)+\partial_tk_s^\infty(z)
   -\partial_sk_t^\infty(z),
\label{fourxeighteen}
\end{equation}
when the lattice-size is {\it infinite}, $L\to\infty$. In this expression,
$k_\mu^\infty(z)$, $k_t^\infty(z)$ and~$k_s^\infty(z)$ are gauge invariant
local currents (which is translational invariant) and the main part~$p(z)$ of
$q^\infty(z)$ is given by\footnote{The numerical coefficient of this expression
cannot be determined by the cohomological analysis. We have used a matching
with a result in the classical continuum
limit~\cite{Luscher:2000un,Adams:2000yi}; see also ref.~\cite{Igarashi:2002zz}
and references therein.}
\begin{eqnarray}
   p(z)&=&{(-1)^{d/2+1}\epsilon_{\rm H}\over2(4\pi)^{d/2}(d/2+1)!}\,
   \epsilon_{M_1N_1\cdots M_{d/2+1}N_{d/2+1}}
   f_{M_1N_1}(z)f_{M_2N_2}(z+\hat M_1+\hat N_1)\cdots
\nonumber\\
   &&\qquad\qquad\qquad\times f_{M_{d/2+1}N_{d/2+1}}
   (z+\hat M_1+\hat N_1+\cdots+\hat M_{d/2}+\hat N_{d/2}),
\label{fourxnineteen}
\end{eqnarray}
where $M=(\mu,t,s)$ etc.\ and we take $\hat t=\hat s=0$;
$f_{r\mu}(z)=u_\mu(z)^{-1}\partial_r u_\mu(z)/i-\partial_\mu a_r(z)$. When
$p(z)$ does not depend on $a_r(z)$,\footnote{For example, when $a_r(z)$ is
pure-gauge $a_r(z)=\omega(z)\partial_r\omega(z)^{-1}/i$, a dependence of~$p(z)$
on~$a_r(z)$ disappears combined with the gauge degrees of freedom~$\omega(z)$
in~eq.~(\ref{threexthirteen}). This is precisely the situation we will consider
below.} one may rewrite $p(z)$ in terms of the reduced gauge field~$U_\mu$ in
an analogous form as eq.~(\ref{threextwelve}). Note that $f_{r\mu}(z)$ is given
by $(T_\mu^0U_\mu^\dagger\partial_r U_\mu T_\mu^{0\dagger})_{m(x)m(x)}/i$.
Then, by the same way as for eq.~(\ref{threextwelve}), one sees that $p(z)$ is
a linear combination of $\str(T^{a_1}\cdots T^{a_{d/2+1}})$.

Now let us evaluate the first Chern number~${\cal I}$~(\ref{fourxten}) by
taking a 2~torus~$T^2$ as the two-dimensional surface~${\cal M}$. We
parameterize $T^2$ by~$0\leq t\leq2\pi$ and~$0\leq s\leq2\pi$. As already
noted, ${\cal I}$ is independent of~$a_\mu^{\rm T}(x)$
in~eq.~(\ref{threexthirteen}); we can set~$a_\mu^{\rm T}(z)=0$ without loss of
generality. Similarly, we may assume that the gauge degrees of
freedom~$\omega(x)$ and the Wilson-line degrees of freedom~$u_\mu^{[w]}(x)$
in~eq.~(\ref{threexthirteen}) have the following standard forms:
\begin{equation}
   \omega(z)=\exp[iL^t(x)t+iL^s(x)s],
\label{fourxtwenty}
\end{equation}
and
\begin{equation}
   u_\mu^{[w]}(z)=\cases{\exp(iJ_\mu^t t+iJ_\mu^s s),&for $x_\mu=0$,\cr
                         1,&otherwise,\cr}
\label{fourxtwentyone}
\end{equation}
where $L^r(x)$ and~$J_\mu^r$ are integer winding numbers, $L^r(x)$,
$J_\mu^r\in{\mathbb Z}$, because these are representatives of the homotopy
class of mappings from~$T^2$ to $\U(1)=S^1$; any mapping can smoothly be
deformed into these standard forms without changing the
integer~${\cal I}$~(\ref{fourxten}).\footnote{When the gauge group is $\SU(N)$,
$\omega_\mu\in{\mathbb Z}_{L^{d-1}}$ and a non-trivial winding of the Wilson
line is impossible. This leads to, as we will see, ${\cal I}=0$
for~${\cal M}=T^2$.} For gauge fields along the continuous directions, we take
the pure gauge configuration, $a_r(z)=\omega(z)\partial_r\omega(z)^{-1}/i%
=-L^r(x)$. Note that this $a_r(z)$ is single-valued on~$T^2$ and thus
eq.~(\ref{fourxseventeen}) holds. Under these restrictions on the gauge fields,
we note
\begin{equation}
   D_r^au_\mu(z)u_\mu(z)^{-1}=iJ_\mu^r\delta_{x_\mu,0},
\label{fourxtwentytwo}
\end{equation}
and
\begin{equation}
   f_{\mu\nu}(z)={2\pi\over L^2}\,m_{\mu\nu},\qquad
   f_{r\mu}(z)=J_\mu^r\delta_{x_\mu,0},\qquad
   f_{ts}(z)=0.
\label{fourxtwentythree}
\end{equation}

For the admissible configuration~(\ref{threexthirteen}) with the above
restrictions on the gauge fields, it is immediate to evaluate the integral
of~$q^\infty(z)$:
\begin{eqnarray}
   &&{1\over2\pi}\int_0^{2\pi}{\rm d}t\,\int_0^{2\pi}{\rm d}s
   \sum_{x\in{\mit\Gamma}}q^\infty(z)
   ={1\over2\pi}\int_0^{2\pi}{\rm d}t\,\int_0^{2\pi}{\rm d}s
   \sum_{x\in{\mit\Gamma}}p(z)
\nonumber\\
   &&={(-1)^{d/2}\epsilon_{\rm H}\over2^{d/2-1}(d/2-1)!}\,
   \epsilon_{\mu_1\nu_1\cdots\mu_{d/2}\nu_{d/2}}
   m_{\mu_1\nu_1}\cdots m_{\mu_{d/2-1}\nu_{d/2-1}}
   J_{\mu_{d/2}}^tJ_{\nu_{d/2}}^s,
\label{fourxtwentyfour}
\end{eqnarray}
which is an integer. The field~$q^\infty(z)$, which is originally defined on
the infinite lattice, depends on the gauge-field background defined on the
infinite lattice. As this gauge-field configuration on the infinite lattice, we
take periodic copies of a gauge-field configuration defined on~${\mit\Gamma}$.
Then, due to the translational invariance, $k_\mu^\infty(z)$ is periodic
on~${\mit\Gamma}$ and we have the first equality. The second equality follows
{}from~eq.~(\ref{fourxtwentythree}).

We can in fact show that (appendix~A), using the locality of the Dirac
operator, integral~(\ref{fourxtwentyfour}) coincides with
eq.~(\ref{fourxseventeen}) when the lattice size~$L$ is sufficiently large,
i.e., when $N$ is sufficiently large. Thus we have
\begin{equation}
   {\cal I}={(-1)^{d/2}\epsilon_{\rm H}\over2^{d/2-1}(d/2-1)!}\,
   \epsilon_{\mu_1\nu_1\cdots\mu_{d/2}\nu_{d/2}}
   m_{\mu_1\nu_1}\cdots m_{\mu_{d/2-1}\nu_{d/2-1}}
   J_{\mu_{d/2}}^tJ_{\nu_{d/2}}^s.
\label{fourxtwentyfive}
\end{equation}
This shows that ${\cal I}\neq0$ for certain configurations defined
on~${\mit\Gamma}\times T^2$ and there exists an obstruction to a smooth measure
on a 2~torus embedded in the space of admissible configurations. As shown
in~section~3, however, the pure-gauge action of any configuration which leads
to ${\cal I}\neq0$ for ${\cal M}=T^2$ diverges as~$N\to\infty$ when~$d>2$,
within the $\U(1)$~embedding.

We want to comment on the difference of our result from Neuberger's
work~\cite{Neuberger:1998xn}. In~ref.~\cite{Neuberger:1998xn}, a torus in the
{\it orbit space}, ${\mathfrak U}/{\mathfrak G}$ where ${\mathfrak U}$ is a
connected component of the space of admissible configurations and
${\mathfrak G}$ is the group of gauge transformations, is considered. It was
then shown that, when the gauge anomaly is not canceled, ${\cal I}\neq0$ for
appropriate configurations. This is an obstruction to define a smooth
${\mathfrak G}$-invariant fermion measure, i.e., an obstruction to the gauge
invariance. See also refs.~\cite{Adams:2000yi,Adams:2002ms}. On the other hand,
we have shown here that there exists an obstruction to a smooth fermion measure
irrespective of its gauge invariance. Even one sacrifices the gauge invariance,
there remains an obstruction.

One might argue that if the gauge invariance is sacrificed, there exists at
least one possible choice of a smooth fermion measure, the Wigner-Brillouin
phase choice~\cite{Narayanan:1993wx}. However, there is a simple example with
which the Wigner-Brillouin phase choice becomes singular, at least with a use
of the overlap Dirac operator (appendix~B). So this choice does not provide
a counter-example for our result.

\section{Conclusion}

In this paper, we systematically investigated possible chiral anomalies in the
reduced model within a framework of the $\U(1)$~embedding. When the
overlap-Dirac operator is employed for the fermion sector, the gauge-field
configuration must be admissible. This admissibility divides the otherwise
connected space of gauge-field configurations into many components. Using the
classification of ref.~\cite{Luscher:1999du}, we gave a general form of the the
axial anomaly~$Q$ within the $\U(1)$~embedding. We have also shown that there
may exist an obstruction to a smooth fermion integration measure in reduced
chiral gauge theories, by evaluating the first Chern number~${\cal I}$ of a
$\U(1)$ bundle associated to the fermion measure. In both cases, the pure gauge
action of gauge-field configurations which cause these non-trivial phenomena
turns to diverge in the 't~Hooft $N\to\infty$ limit when~$d>2$. This might
imply that the above phenomena are irrelevant in the 't~Hooft $N\to\infty$
limit, in which the reduced model is considered to be equivalent to the
original gauge theory.

The most important question we did not answer in this paper is an effect of the
$\U(1)$ embedding to other gauge representations. This is related to a question
of the gauge anomaly cancellation in reduced chiral gauge theories. We expect
that if the fermion multiplet is anomaly-free in the conventional sense, then
the obstruction we found in the reduced model will disappear. To see this,
however, we have to evaluate~${\cal I}$ for a Weyl fermion belonging to a
representation~$R$, with the gauge-field configuration\footnote{For the
``trivial'' anomaly-free cases which consist of equal number of right-handed
and left-handed Weyl fermions in the fundamental representation, the
obstruction~${\cal I}$ vanishes because ${\cal I}$ is proportional to the
chirality~$\epsilon_{\rm H}$.}
\begin{equation}
   R(u_\mu T_\mu^0).
\label{fivexone}
\end{equation}
Of course, it may be possible to imitate the $\U(1)$ embedding in other
representations by restricting gauge-field configurations as
\begin{equation}
   R(U_\mu)=u_\mu'T_\mu^{0\prime},
\label{fivextwo}
\end{equation}
where $R$ is a $N'\times N'$ representation matrix and the shift operator
$T_\mu^{0\prime}$ is for a lattice with the size $L'$ and $L^{\prime d}=N'$.
A similar argument as this paper will then be applied with this type of
embedding. Generally, however, the backgrounds~(\ref{fivexone})
and~(\ref{fivextwo}) do not coincide. For the case of the adjoint
representation, a connection of the reduced model to non-commutative lattice
gauge theory~\cite{Ambjorn:1999ts,Nishimura:2001dq} might be helpful.

Another interesting extension is to embed a lattice gauge theory with a larger
gauge group, say $\SU(2)$, in the reduced model. This is easily done at least
for the fundamental representation by identifying two or more columns of the
representation vector as a single lattice site. A freedom of internal space
then emerges. With this embedding, we have to analyze the axial anomaly in
non-abelian lattice gauge theories defined on a finite-size lattice. As for the
corresponding axial anomaly~$Q$, there is a conjecture~\cite{Igarashi:2002zz},
which holds to all orders in perturbation theory, that $Q$ coincides with the
L\"uscher's topological charge~\cite{Luscher:1981zq}. So, accepting this
conjecture, the $\SU(2)$ instanton configuration on the
lattice~\cite{Laursen:ec} with this embedding will provide an example
of~$Q\neq0$.

Another direction is to investigate the Witten anomaly~\cite{Witten:1982fp} in
the present setup following the line of argument in
refs.~\cite{Neuberger:1998rn,Bar:2000qa}.

So, there are many things to do with this embedding trick in the reduced model,
when a Ginsparg-Wilson type Dirac operator is employed. We hope to come back
some of above problems in the near future.

The authors would like to thank Jun Nishimura for valuable discussions. We
would like to thank David Adams for pointing out a misleading statement in the
first version of this paper. H.S. would like to thank Kiyoshi Okuyama and
Kazuya Shimada for helpful discussions on the reduced model. This work is
supported in part by Grant-in-Aid for Scientific Research, \#12640262,
\#14046207 (Y.K.) and \#13740142 (H.S.).

\appendix

\section{Proof of eq.~(\ref{fourxtwentyfive})%
\protect\footnote{A part of this proof was obtained through H.S.'s discussion
with Takanori Fujiwara and Keiichi Nagao.}}

When the size of~${\mit{\Gamma}}$ becomes infinity, $L\to\infty$, a
Ginsparg-Wilson Dirac operator $D(x,y)$ is promoted to a Dirac operator on the
infinite lattice $D(x,y)\to D^\infty(x,y)$. We assume that these two operators
are related by the reflection~\cite{Luscher:1999du}
\begin{equation}
   D(x,y)=\sum_{n\in{\mathbb Z}^d}D^\infty(x,y+Ln),
\label{axone}
\end{equation}
where the gauge field configuration in the right hand side is given by periodic
copies of~${\mit{\Gamma}}$ extended to the infinite lattice. This relation
actually holds for the overlap-Dirac operator. Equation~(\ref{axone}) implies,
when $a_r(z)$ is pure-gauge,
\begin{eqnarray}
   &&\sum_{x\in{\mit\Gamma}}q(z)
   =i\sum_{x\in{\mit\Gamma}}
   \tr(\hat P_{\rm H}[D_t^a\hat P_{\rm H},D^a_s\hat P_{\rm H}])(x,x)
\nonumber\\
   &&=i\sum_{n\in{\mathbb Z}^d}\sum_{x\in{\mit\Gamma}}
   \sum_{y,z\in{\mathbb Z}^d}
   \tr\{\hat P_{\rm H}^\infty(x,y)
   [D_t^a\hat P_{\rm H}^\infty(y,z)D_s^a\hat P_{\rm H}^\infty(z,x+Ln)
   -(t\leftrightarrow s)]\},
\label{axtwo}
\end{eqnarray}
where the kernel $\hat P_{\rm H}^\infty(x,y)$ is defined from $D^\infty(x,y)$.
Note that a sum of~$q^\infty(z)$ over~${\mit\Gamma}$ in
eq.~(\ref{fourxtwentyfour}), $\sum_{x\in{\mit\Gamma}}q^\infty(z)$, coincides
with the $n=0$~term of~eq.~(\ref{axtwo}). On the other hand, from the locality
of the Dirac operator (see ref.~\cite{Luscher:1999du}), it is possible to show
bounds
\begin{eqnarray}
   &&\|\hat P_{\rm H}^\infty(x,y)\|\leq
   \kappa_1(1+\|x-y\|^{\nu_1})e^{-\|x-y\|/\varrho},
\nonumber\\
   &&\|D_r^a\hat P_{\rm H}^\infty(x,y)\|\leq\kappa_2(1+\|x-y\|^{\nu_2})
   e^{-\|x-y\|/\varrho}\max_{x,\mu}|D_r^au_\mu(z)u_\mu(z)^{-1}|,
\label{axthree}
\end{eqnarray}
where the constants~$\kappa_1$, $\kappa_2$, $\nu_1$ and~$\nu_2$ are independent
of the gauge field. We thus have the bound
\begin{equation}
   \biggl|\sum_{x\in{\mit\Gamma}}q(z)
   -\sum_{x\in{\mit\Gamma}}q^\infty(z)\biggr|
   \leq\kappa_3L^{\nu_3}e^{-L/\varrho}
   \max_\mu|J_\mu^t|\max_\nu|J_\nu^s|,
\label{axfour}
\end{equation}
where a use of eq.~(\ref{fourxtwentytwo}) has been made. This shows
\begin{equation}
   \biggl|\,{\cal I}-{1\over2\pi}\int_0^{2\pi}{\rm d}t\,\int_0^{2\pi}{\rm d}s
   \sum_{x\in{\mit\Gamma}}q^\infty(z)\biggr|
   \leq\kappa_4L^{\nu_4}e^{-L/\varrho}
   \max_\mu|J_\mu^t|\max_\nu|J_\nu^s|,
\label{axfive}
\end{equation}
and, when the lattice size is sufficiently large, say $L/\varrho>n$, the
integer~${\cal I}$ and the integer~(\ref{fourxtwentyfour}) coincide. The
required lattice-size for this coincidence however may depend on the
gauge-field configuration through the winding numbers~$J_\mu^r$.

\section{Wigner-Brillouin phase choice may become singular\protect\footnote{The
following example was suggested to us by Martin L\"uscher in the context of
general lattice chiral gauge theories.}}

Consider a one-parameter family of gauge-field configurations in
$\U(1)$~theory:
\begin{equation}
   u_\mu^{(\tau)}(x)=\cases{e^{i\pi\tau},&for $\mu=1$,\cr
                           1,&otherwise,\cr}
\label{bxone}
\end{equation}
where $0\leq\tau\leq1$. The field strength of these configurations vanishes,
$f_{\mu\nu}^{(\tau)}(x)=0$, so these are admissible configurations. The
modified chiral matrix and the projection operator corresponding to these
configurations will be denoted by~$\hat\gamma_{d+1}^{(\tau)}$
and~$\hat P_{\rm H}^{(\tau)}$. From the definition of the overlap-Dirac
operator, one then finds
\begin{equation}
   \hat\gamma_{d+1}^{(\tau)}\psi
   =\gamma_{d+1}(-i\gamma_1\sin\pi\tau+\cos\pi\tau)\psi,
\label{bxtwo}
\end{equation}
for any {\it constant\/} spinor~$\psi$. This implies
\begin{equation}
   \hat P_{\rm H}^{(1)}\psi=\hat P_{\widetilde H}^{(0)}\psi.
\label{bxthree}
\end{equation}

Now, in the Wigner-Brillouin phase choice, the phase ambiguity of the fermion
measure is fixed by imposing $\det(v_j^{(0)},v_k^{(\tau)})$ be real positive,
where basis vectors satisfy $\hat P_{\rm H}^{(0)}v_j^{(0)}=v_j^{(0)}$
and~$\hat P_{\rm H}^{(\tau)}v_j^{(\tau)}=v_j^{(\tau)}$. This determinant,
however, vanishes at~$\tau=1$ because $\hat P_H^{(1)}\psi$ is contained
in~$v_j^{(1)}$ and
\begin{equation}
   (v_j^{(0)},\hat P_{\rm H}^{(1)}\psi)
   =(v_j^{(0)},\hat P_{\widetilde{\rm H}}^{(0)}\psi)=0.
\label{bxfour}
\end{equation}
Therefore the Wigner-Brillouin phase choice becomes singular at~$\tau=1$.

\listoftables           
\listoffigures          

\end{document}